# Structural & "Granger" CAUSALITY for IoT Digital Twin

Dr. PG Madhavan

Seattle, USA March 2022
pg@jininnovation.com

#IoT #Multichannel #Structuralcausality #Grangercausality #Digitaltwin #Causaldigitaltwin #Fencegraph #Causality #Simulation

**Abstract**: *In this foundational expository article on the application of Causality Analysis in IoT, we establish the basic theory and algorithms for estimating Structural and Granger causality factors from measured multichannel sensor data (vector timeseries). Vector timeseries is modeled as a Structural Vector Autoregressive (SVAR) model; utilizing Kalman Filter and Independent Component Analysis (ICA) methods, Structural and generalized Granger causality factors are estimated. The estimated causal factors are presented as a "Fence" graph which we call "Causal Digital Twin". Practical applications of Causal Digital Twin are demonstrated on NASA Prognostic Data Repository Bearing data collection. Use of Causal Digital Twin for "counterfactual" experiments are indicated.*

*Causal Digital Twin is a horizontal solution that applies to diverse use cases in multiple industries such as Industrial, Manufacturing, Automotive, Consumer, Building and Smart City.*

In this report, we exposit the methods and use of Causality Analysis for extracting non-obvious information from multichannel sensor data in IoT use cases.

As a succinct summary, "*IoT is about instrumenting things, connecting them, and applying analytics to extract meaning and insights from the data harvested. Digital twin is a digital reflection of the physical state and condition of a unique physical thing. Digital twins are used to give digital meaning to data that represent the dynamic attributes of a thing that you are monitoring. Simulation applications built on top of a digital twin (and its data exhaust or historical data)*" - they can provide prescriptive insights that can not only save expenses but enhance performance – this requires knowledge of cause-effect relationships among "things"; such a solution is called "Causal" digital twin (CDT). In the absence of Causal information, we can only treat the symptoms; without knowing root-causes, systematic improvement in performance of a machine or a factory or a city is impossible to achieve.

IoT collects data from multiple sensors continuously and taken together, they form a "vector" time series. At a minimum, the model of this multichannel data should capture the auto-correlation and cross-correlation structure of this vector time series and estimate model parameters of a Structural Vector Autoregressive (SVAR) model (or more complicated stochastic models**). Model parameters can then be interpreted as Structural Causality and Granger Causality. We use Kalman Filter and Independent Component Analysis (ICA) methods to estimate SVAR parameters.**

One of the main discoverer and proponent of ICA is Aapo Hyvarinen; this is the fundamental source: Hyvarinen, et al., "Estimation of a Structural Vector Autoregression Model Using Non-Gaussianity", Journal of Machine Learning Research, 2010.

My own article, "Causality & Counterfactuals: Role in IoT Digital Twins" (2021), provides explanations in the IoT context.

## SVAR Model & Causality

We model multichannel IoT time series as a Structural Vector Autoregressive (SVAR) model. In the **Appendix ("Regression coefficients ARE Causal Factors")**, we explain why SVAR model parameters are Causal Factors.

For a multichannel (vector) timeseries, $\mathbf{y}_t$ -

$$\mathbf{y}_t = \mathbf{S}^0 \mathbf{y}_t + \sum_{d=1}^{D} \mathbf{S}^d \mathbf{y}_{t-d} + \mathbf{e}_t \quad \ldots (1)$$

$\mathbf{S}^0$ is the "Structural" causal factor matrix whose diagonal entries are ZERO, $\mathbf{S}^d$ are the "Lagged" causal factors and $\mathbf{e}$ is the model error.

Consider only a few terms to understand the discrete-time SVAR equation better. Further, consider a simpler case of only 2 channels and 1 lag. The corresponding equations are –

$y_1[n] = 0 * y_1[n] + s_{12}^0 y_2[n] + s_{11}^1 y_1[n-1] + s_{12}^1 y_2[n-1] + e_1[n] \quad \ldots (2)$

$y_2[n] = s_{21}^0 y_1[n] + 0 * y_2[n] + s_{21}^1 y_1[n-1] + s_{22}^1 y_2[n-1] + e_2[n] \quad \ldots (3)$

The matrix elements, "s", are the "regression" coefficients in equations (2) and (3) above. Color coding is such that Structural causal factors are BLUE and Lagged causal factors are GREEN (*we will discuss why they are the causal factors presently*).

A diagram corresponding to equations (2) and (3) is shown below (without error terms).

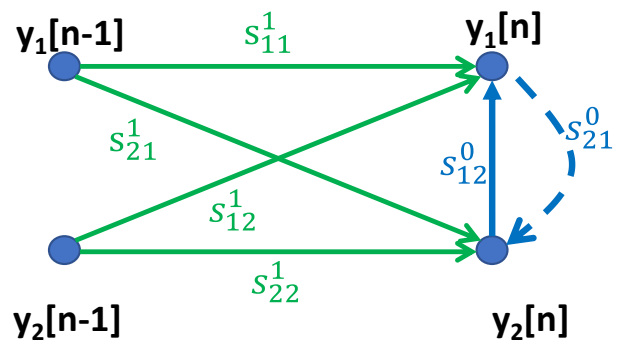

Figure 1. Causal graph of simplified SVAR model.

Considering the current time index, 'n', and equation (2) and (3),

- Structural causal factors from $y_2[n]$ (solid BLUE arrow top) and Lagged causal factors from $y_1[n-1]$ and $y_2[n-1]$ (2 GREEN arrows terminating at $y_1[n]$) affect $y_1[n]$.

- For $y_2[n]$, Lagged causal factors from $y_1[n-1]$ and $y_2[n-1]$ (2 GREEN arrows terminating at $y_2[n]$) affect $y_2[n]$.
- $y_1[n]$ affecting itself is not allowed – no "self" causality.
- The dotted BLUE arrow from $y_1[n]$ to $y_2[n]$ will be discussed in the Appendix.

## Granger Causality

Traditional Granger causality between two time series is defined as the ratio of prediction error variance of one time series over the reduced prediction error variance if the lagged elements of the OTHER time series is also brought into play.

Consider a 2-channel time series, i.e., $[y_1(n)\ y_2(n)]^T$

Assuming AR(1) processes, the following prediction errors can be defined following equation 1.

$e_t = y_{1t} - \sum_{d=1}^{D} s1^d\, y_{1t-d}$

$\varepsilon_t = y_{1t} - \sum_{d=1}^{D} s1^d\, y_{1t-d} - \sum_{d=1}^{D} s2^d\, y_{2t-d}$

If $\text{var}[\varepsilon_t] < \text{var}[e_t]$, $y_2$ "Granger-causes" $y_1$.

Granger Causality, $\mathcal{F}_{(y_2 \to y_1)} = \ln \frac{\text{var}[e_t]}{\text{var}[\varepsilon_t]}$

NOTE: Traditional Granger causality is defined ONLY for 2-channel time series; there are extensions but SVAR model does it nicely. Hyvarinen, et al., explains why $\mathbf{S}^d$ in equation 1 is a generalization of Granger Causality (though not in the variance ratio form). Note that diagonal elements of $\mathbf{S}^d$ are ignored when using SVAR for Granger causality (because Granger causality accounts for *cross-channel* prediction ability).

## Multichannel IoT Causal ("MIC") digital twin:

When we put the pieces together for a multichannel (vector) time series using SVAR model, we get "Structural Causal" factor estimates ($= \mathbf{S}^0$) and "Generalized" Granger Causal factor estimates (= non-diagonal elements of $\mathbf{S}^d$).

Various algorithms exist for SVAR estimation. When $\mathbf{S}^0$ is *triangular*, *DAG restriction is satisfied which makes SVAR parameter estimates "CAUSAL Factors"*. Independent Component Analysis (ICA) algorithm for Blind Signal Separation problem ($\mathbf{S}^0$) solution, when model noise is non-Gaussian & Independent under LiNGAM model (Linear, Non-Gaussian, Acyclic Model), is used as our estimation method.

*(SVAR, ICA, LiNGAM, Blind Signal Separation (sometimes known as "cocktail party" problem) have a celebrated history and significant theoretical underpinnings. There are fast (even fixed-point) algorithms, free Matlab and Python code and other support widely available. This work has matured nicely over the last 20 years.)*

## Estimating SVAR coefficients

Reproducing SVAR model in equation 1 below,

$\mathbf{y_t} = \mathbf{S^0 y_t} + \sum_{d=1}^{D} \mathbf{S^d\, y_{t-d}} + \mathbf{e_t}$

"Structural Causal" factor estimates ($= \mathbf{S}^0$) and "Multichannel Generalized" Granger Causal factor estimates are obtained by the following steps.

Step 1: VAR Estimation (Kalman Filter)
$y_t = \sum_{d=1}^{D} \mathbf{M^d\, y_{t-d}} + \mathbf{n_t}$ → Estimate $\widehat{\mathbf{M}}^{1..D}$

One type of "Generalized" Granger Causal factor estimate is $\widehat{\mathbf{M}}^d$ (non-diagonal elements) which is a version of $\mathbf{S}^d$ that accords with the definition of Granger causality.

Step 2: Get Residuals
$\hat{\mathbf{n}}_t = \mathbf{y}_t - \sum_{d=1}^{D} \hat{\mathbf{M}}^d \mathbf{y}_{t-d}$ → *Calculate* $\hat{\mathbf{n}}_t$

Step 3: SCM Estimation (using ICA)
$\hat{\mathbf{n}}_t = \mathbf{S}^0 \hat{\mathbf{n}}_t + \mathbf{e}_t$ → Estimate $\hat{\mathbf{S}}^0$
"Structural Causal" factor estimate is $\hat{\mathbf{S}}^0$.

Step 4: Corrected VAR
*For d = 1...D,*
$\hat{\mathbf{S}}^d = (\mathbf{I} - \hat{\mathbf{S}}^0) \hat{\mathbf{M}}^d$ → Calculate $\hat{\mathbf{S}}^{1..D}$

"Corrected" Generalized Granger ("G-Granger") causal factor estimate is the non-diagonal elements of $\hat{\mathbf{S}}^d$. Hyvarinen, et al., (2010) show with simple examples how if one does not "back out" the effect of Structural causality on G-Granger causality, G-Granger causality estimates ($\hat{\mathbf{M}}^d$) may be corrupted. Corrected G-Granger, non-diagonal elements of $\hat{\mathbf{S}}$, avoids this potential error.

# Structural & Multichannel Granger Causality tests with REAL data: NASA Bearing data

We apply the algorithms from the last section and develop a Causal Digital Twin (CDT) in a real-life setting using the popular NASA Prognostics Data Repository's bearing dataset. The data is from a run-to-failure test setup of bearings installed on a shaft. The arrangement is shown in figure 2.

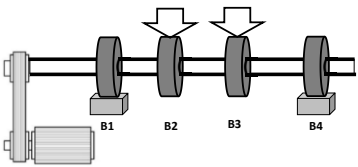

Figure 2. NASA Prognostic Data Repository Bearing data collection setup

4-channel vibration data from each bearing were collected from Feb 12 to Feb 19 and the tests were run continuously to failure and data collected through the entire period. **Bearing #1 's outer race failed on Feb 19.** The 4-channel time-series data of vibration measurements from each of the 4 bearings were subjected to our SVAR model estimation to obtain Structural and Granger causal factors among these 4 bearings.

The results from Feb 19 when Bearing #1 failed can be graphically shown in multiple ways. Considering only lag=1 (D=1 in the SVAR equation), **the "Fence" graph is shown in figure 3. This is what we are calling "Multichannel Causal Digital Twin".**

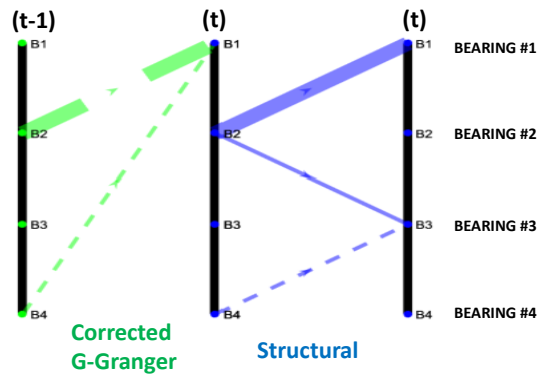

Figure 3. Fence graph on Feb 19

The 4 bearings are labeled vertically on the right-hand side. On the top, (t) indicates the current instant of time and (t-1) indicates Lag =1. There are two (t)'s to show Structural and Granger (defined for lagged cause-effect) in one picture. Thickness of the lines indicate the relative magnitude of the causality factor (BLUE for Structural and GREEN for Granger); solids lines are positive causal factors (drives the effect up) and dotted lines are negative causal factors (drives the effect down).

Causal information can also be shown in more traditional ways as shown below. In one case, you can only see the Structural factors and in the other, lag =1 is shown but more lags will be hard to visualize.

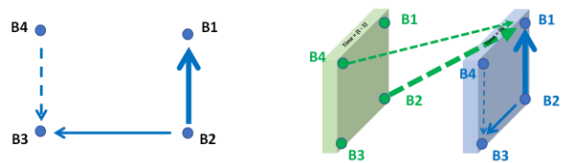

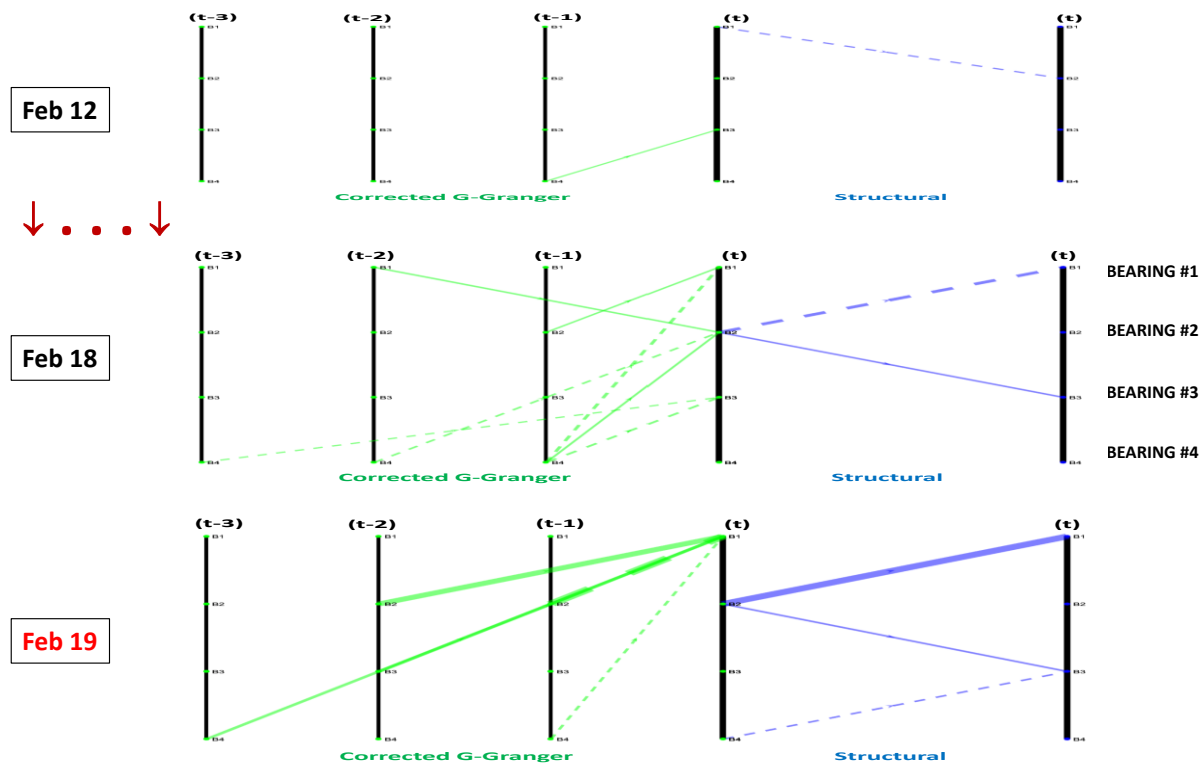

Figure 4. Progression of Multichannel Causal Digital Twin (Structural and Granger causality) over days

It turns out that the 4 vibration timeseries have individual models larger than AR(1). So we estimated Granger causality up to Lag=3.

Note that all GREEN (Granger) links terminate on the vertical (t), 2nd from the right. In other words, Granger causality only considers how the current instant of one timeseries is affected by PAST instances of OTHER timeseries.

## Value of Information from Causal Factors

We note the following important insights.

- *Multichannel Causal Digital Twin (MCDT) is NOT for abnormality detection!* There are many simple methods (such as thresholding the bearing vibration amplitude) but MCDT may also be used for more sensitive and robust detection.

Comparing the results for various days, it is obvious that additional lags have a lot to say about machine dynamics.

- From figure 4, complexity of Structural causality and Granger causality patterns increase over the days of testing. Machine dynamics meaning of this is that more and more vibration energy is being coupled among the bearings beyond what was in the original design. Certain couplings can be detrimental to the RUL (remaining useful life) of the bearings.
- In this NASA test data, it was Bearing #1 that failed on Feb 19. On Feb 19, you can see that Bearing #2 is strongly coupled to Bearing #1 meaning that the former is pouring vibration energy into the latter.
- "Energy coupling" does not happen only at the current instant, (t)! A previous instant can couple energy into Bearing #1 at the current instant and thus set up something like a

- "standing wave" oscillation which will reduce RUL.
- Interaction between Bearing #4 and #1: Bearing #1 has a positive (Granger) causal factor from 3 instances ago (lag=3). Such a delayed interaction (perhaps precipitated by the amount of increasing shaft misalignment), can set up oscillations that may flex the whole experimental setup!
- Granger causality pattern gets more complex across days, even before Structural causality. This may be a very powerful prediction mechanism in that lagged causality is an earlier *precursor* of failure than structural causality.

As we have shown in a previous article, Madhavan (Aug 2021), **the real power of Multichannel CDT using SVAR is the ability to perform "counterfactual" experiments** whereby simulating scenarios using MCDT, we can perform what-if analysis on data that we have in hand (rather than having to execute expensive new data collection) to find ways to improve performance and even design.

### References

Hyvarinen, A, et al., "Estimation of a Structural Vector Autoregression Model Using Non-Gaussianity", Journal of Machine Learning Research, 2010.

Madhavan, PG,, "Causality & Counterfactuals: Role in IoT Digital Twins", Oct 2021.

Madhavan, PG, "Multichannel IoT Causal (MIC) digital twin: Counterfactual experiments on Fence Graphs", Aug 2021.

### About the author:

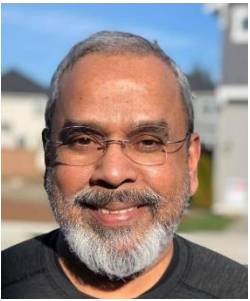

After obtaining his Ph.D. in Electrical and Computer Engineering from McMaster University, Canada, and Masters in Biomedical Engineering from IIT, Madras, Dr. Madhavan pursued original research in Random Field Theory and Computational Neuroscience as a professor at the University of Michigan, Ann Arbor, and Waterloo University, Canada, among others. His next career in corporate technology saw him assume product leadership roles at Microsoft, Bell Labs, Rockwell Automation, GE Aviation and lastly at NEC Corporation, Japan. Later, PG founded and was CEO at 2 startups (and CTO at 2 others) leading all aspects of startup life.

PG launched his first IoT product at Rockwell Automation back in 2000 for predictive maintenance, an end-to-end solution including a display digital twin. Since then, he has been involved in the development of IoT technologies such as fault detection in jet engines at GE Aviation and causal digital twins to improve operational outcomes.

His collected data science works in IoT has been published as a book, "Data Science for IoT Engineers" in Jan 2022.

PG has 12 issued US Patents and over 50 published articles.

# APPENDIX: Regression coefficients ARE Causal Factors

"Causation is MORE THAN Correlation". To investigate, consider a single channel, total number of lags, D=1, case of the SVAR model in equation (1).

$$y_1[n] = s_{11}^1 \, y_1[n-1] + e[n]$$

Clearly, this is a simple linear regression equation with $s_{11}^1$ as the regression coefficient. You will also recognize it as an autoregressive process, AR(1), with $s_{11}^1$ as the AR(1) coefficient. From basic Time Series Analysis course work, you know that the estimate of AR(1) coefficient is the lag=1 autocorrelation, i.e., correlation between $y_1[n]$ and $y_1[n-1]$.

But in the SVAR model, we have called it "Lagged causal factor"! *So, is causal factor just a correlation coefficient*???

Let us take the next step of studying the single channel but D=2 case of SVAR model equation.

$$y_1[n] = s_{11}^1 \, y_1[n-1] + s_{11}^2 \, y_1[n-2] + e[n]$$

This is an AR(2) process with $s_{11}^1$ and $s_{11}^2$ as AR coefficients which are estimated from the lag 1 and 2 correlations via the solution of Yule-Walker equations and they are –

$$s_{11}^1 = \frac{\rho_1(1-\rho_2)}{1-\rho_1^2}, \quad s_{11}^2 = \frac{\rho_2 - \rho_1^2}{1-\rho_1^2}; \quad \rho_{1,2} - \text{ Autocorrelation lag 1 \& 2}$$

So, the SVAR Causal factors, $s_{11}^1, s_{11}^2$, are **not pure Correlations** anymore! In general, they will something MORE than correlations.

As we observed above, "Causation is Correlation *PLUS something else*." It turns out that if we make certain assumptions, SVAR coefficients (which are "more than" correlations) can be interpreted as "Causal Factors".

Assumptions that elevate certain Correlations to CAUSATION:
  Acyclic: No cycles in the corresponding graph
  Markov: All associations between variables are due to causal relations
  Faithfulness: Causal influence is not hidden by coincidental cancelations

These are from the foundational works of Judea Pearl, Peter Spirtes and others in the last two decades. What they do is restrict the graph in figure 1 to a "DAG" – directed acyclic graph. DAG assumption does not allow self-causality (which is logical). Since cycles are not allowed in a DAG, only one of $s^0{}_{12}$ or $s^0{}_{21}$ can be non-zero (otherwise, you cannot decide which causes which).

In conclusion, ***based on DAG assumption***,
**$S^0$ and $S^d$ CAUSAL factors are multiple linear regression coefficients of SVAR model.**

So, regression coefficients ARE indeed causal factors under the assumptions above . . .